\documentclass[twocolumn]{aastex631}

\shorttitle{Neutrino Echos}
\shortauthors{Gullin et al.}
\graphicspath{{./}{figures/}}

\begin{document}

\title{Neutrino Echos following Black Hole Formation in Core-Collapse Supernovae}

\author{Samuel Gullin}
\author[0000-0002-8228-796X]{Evan P. O'Connor}
\affiliation{The Oskar Klein Centre, Department of Astronomy,\\ Stockholm University, AlbaNova, SE-106 91 Stockholm, Sweden}

\author{Jia-Shian Wang}
\author[0000-0003-1731-5853]{Jeff Tseng}
\affiliation{%
Department of Physics, Oxford University, Oxford OX1 3RH United Kingdom
}%

\begin{abstract}

During a failed core-collapse supernova, the protoneutron star eventually collapses under its own gravitational field and forms a black hole.  This collapse happens quickly, on the dynamical time of the protoneutron star, $\lesssim$0.5\,ms.  During this collapse, barring any excessive rotation, the entire protoneutron star is accreted into the newly formed black hole.  The main source of neutrinos is now removed and the signal abruptly shuts off over this formation timescale. However, while the source of neutrinos is turned off, the arrival times at an Earth-based detector will depend on the neutrino path.  We show here that a modest amount of neutrinos, emitted just prior to the black hole forming, scatter on the infalling material into our line of sight and arrive after the formation of the black hole, up to 15\,ms in our model.  This neutrino echo, which we characterize with Monte Carlo simulations and analytic models, has a significantly higher average energy (upwards of $\sim$ 50\,MeV) compared to the main neutrino signal, and for the canonical failed supernova explored here, is likely detectable in $\mathcal{O}$(10\,kT) supernova neutrino detectors for Galactic failed supernovae.  The presence of this signal is important to consider if using black hole formation as a time post for triangulation or the post black hole timing profile for neutrino mass measurements. On its own, it can also be used to characterize or constrain the structure and nature of the accretion flow.

\end{abstract}

\section{Introduction} \label{sec:intro}

Core-Collapse Supernovae (CCSNe) are true multi-messenger events.  The extreme densities, temperatures, and spacetime curvatures produce neutrinos, gravitational waves, and in many cases, explosive events that give rise to brilliant electromagnetic displays throughout the Universe.  CCSNe are also the birth site of neutron stars and black holes.  The latter have several formation channels, including in failed CCSNe \citep{oconnor:11}, in successful CCSNe with substantial fallback \citep{zhang:08}, or in an intermediate regime, so-called aborted CCSN \citep{chan:18}.  In all three of these cases, the protoneutron star initially formed after the core collapse accretes too much mass and cannot support itself against gravitational collapse to a black hole. In failed and aborted supernovae, the black hole typically forms on relatively short timescales ($\sim 1\,$s) following the core collapse.  At these times, the neutrino emission is still strong, and an imprint of the black hole formation process in the neutrino signal is expected \citep{burrows:88,baumgarte:96,beacom:01,li:21}, with a relatively abrupt (compared to the total duration of the neutrino signal) shut off. For failed CCSN in the Galaxy, neutrinos will be the first messenger of this extraordinary event and neutrino detectors are ready to observe and alert observers \citep{snews2p0}.  This neutrino emission, particularly the emission near and following the black hole formation is the main focus of this work.

Failed CCSNe begin like normal CCSNe \cite{oconnor:11}. Following the core collapse, when the matter reaches nuclear density, the equation of state stiffens and the core undergoes a bounce, launching the supernova shock wave and starting the intense neutrino emission phase. In garden-variety successful supernovae, this bounce shock is revived by the explosion mechanism, which most theoretical evidence from state-of-the-art, three dimensional simulations suggest is the neutrino mechanism \citep{bethe:85, burrows:20, bollig:21}.  Following the explosion, there can be continued accretion, which as shown in \cite{bollig:21} may be critical to obtaining canonical explosion energies. However, this is likely only a small amount of mass for most successful supernovae, and a neutron star is left as the remnant. As alluded to above, in some cases there could be either substantial fallback that eventually pushes the protoneutron star over the limit and causes collapse to a black hole, or the explosion could happen late enough for a given progenitor that the protoneutron star is already close to or over this limit and therefore will collapse, potentially quickly, to a black hole \citep{chan:18}.  The latter, aborted supernovae, are receiving increasing attention in the literature, following several multidimensional models of extreme stars that produce runaway shock expansion \citep{ott:18,chan:18,pan:18,pan:21,obergaulinger:17,walk:20}, and in the case of \cite{pan:18,walk:20,chan:18,obergaulinger:17}, later produce black holes. For truly failed supernovae, the shock is never revived and accretion onto the protoneutron star continues unabated until its becomes unstable and collapses.  As long as the shock fails to be revived, the timing of this ultimate collapse is set by the progenitor structure \citep{oconnor:11}, the nuclear equation of state and its thermal response \citep{schneider:20}, and to a lesser extent by the neutrino interactions, rotation, and multidimensional dynamics. Failed CCSNe occur in an unknown fraction of core-collapse events. However, long-term observational campaigns monitoring for the disappearance of nearby stars \citep{kochanek:08,gerke:15,adams:17,neustadt:21,villarroel:20}, a prediction of failed supernovae, are providing constraints.  Failed supernova candidates have been identified, and with the latest findings of \cite{neustadt:21} predicting a failed supernova fraction of $f = 0.16^{+0.23}_{-0.12}$ (at 90\% confidence). Theoretical studies of the progenitor dependence of the explosion mechanism with models, while highly parameterized and not without their own caveats, also suggest similar failed CCSN fractions \cite{ertl:16,muller:16,couch:20}. 

The main neutrino signal from a failed CCSN is from the emission of neutrinos from the dense ($\rho \gtrsim 10^{11}$\,g\,cm$^{-3}$) and hot ($T\gtrsim$\,few\,MeV) matter in the protoneutron star that forms after the core collapses. For failed supernovae that quickly, $t \lesssim 1$\,s, form a black hole, a significant portion of the gravitational binding energy released during the collapse is still trapped in the protoneutron star and is accreted into the black hole and not released as neutrinos, nevertheless, the intense accretion rate and high matter temperatures still make these events prodigious producers of neutrinos, with total event rates in Earth-based neutrino detectors in this 1\,s window rivaling the total signal expected from standard successful supernovae \citep{snews2p0,walk:20}.  For protoneutron stars that take a long time to collapse to black holes ($t \gg 1\,$s), for example, in failed core-collapse supernovae in low-compactness progenitors with small accretion rates \citep{kresse:21}, the majority of the released binding energy can escape and the integrated neutrino emission can be very high.  This is further increased by the large binding energy of massive protoneutron stars.  In both cases, the neutrinos emitted during a failed core-collapse supernovae carry important information on the thermodynamics and the structure of the protoneutron star. For this reason, they provide the means to constrain the accretion history, progenitor structure, and the high density equation of state \citep{oconnor:13,suwa:16,kresse:21,schneider:20,walk:20}. 

The study of the neutrino signal near black hole formation from a then-state-of-the-art model was first described in detail in \cite{baumgarte:96}.  Using a neutrino leakage scheme and a spherically-symmetric general-relativistic Lagrangian hydrodynamics code based on \cite{bst:95}, the neutrino luminosity and mean energy were determined for the collapse of a hot protoneutron star to a black hole following the hyperonization of the cooling core. The singularity avoiding code did not permit the presence of an event horizon, but allowed evolution of the matter as it approached the horizon.  At the formation, on the dynamical timescale of the black hole ($\lesssim 0.5\,$ms), and contrary to the often-used language of "abruptly turns off", the neutrino luminosity and average energy begin to decay, eventually approaching an exponential decay as the last neutrinos escaping the protoneutron star are heavily redshifted. The analytic prediction of such a signal, with $L\propto \exp{-t/\sqrt{27}M}$ goes back even farther, although in the context of photons, not neutrinos \citep{podurets:64,ames:68}. This late time luminosity is understood as leakage from a photon cloud near $r=3M$, the photon ring, where a small fraction of photons emitted during the collapse can circle the photon ring many times before escaping to an observer.  In a companion work, \cite{wang:21}, we explore this analytic prediction for neutrinos around stationary and rotating black holes.

In this work we use GR1D \citep{oconnor:10,oconnor:15} to evolve a $40\,M_\odot$, solar-metallicity progenitor from \cite{woosley:07} through core collapse and until the point of black hole formation.  Like \cite{baumgarte:96}, GR1D contains a general relativistic hydrodynamics prescription for modeling the spacetime and matter evolution (although via an Eulerian based approach rather than a Lagrangian). In addition, GR1D uses modern energy dependent neutrino transport algorithms, neutrino interaction rates, progenitors, and an equation of state that can support $2\,M_\odot$ neutron stars.  For our detailed neutrino signal predictions during and following the black hole formation, we take the results from GR1D and use SedonuGR \citep{richers:15,richers:17} a general-relativistic Monte Carlo neutrino transport code, which for enough particles, gives the exact solution of the Boltzmann equation for our setup. We modify SedonuGR to be semi-time-dependent to capture the rapidly changing spacetime and matter distribution near black hole formation. SedonuGR allows for non-radial neutrino propagation, an aspect missing from the work of \cite{baumgarte:96} but crucial to obtain the correct neutrino time profiles \citep{podurets:64,ames:68,wang:21}.  It also can include neutrino processes that cannot be fully modeled with neutrino leakage (or even with the approximate transport methods in GR1D), such as non-local neutrino scattering in the accretion flow. This brings us to the main focus of our work.

In this study, we discover and characterize a new component of the neutrino signal in failed core-collapse supernovae: a neutrino echo which follows black hole formation.  The echo is from neutrinos emitted before the collapse to a black hole that coherently scatter off of nuclei in the supersonic accretion flow surrounding the collapsing protoneutron star (out to about 5000\,km). This increases the path the neutrino take to Earth by up to $\sim$5000\,km/$c \sim 15$\,ms   The neutrino echo provides a direct probe of the properties of this accretion flow.  With the help of an analytic model, we show the characteristic features of this echo, namely that the time dependence following the black hole formation follows from the radial dependence of the matter surrounding the compact object core.  The normalization of the neutrino echo is set by the density and composition of this material, as well as by the spectral properties of the neutrinos emitted before the collapse. Since the latter will be well constrained by the observations of the main signal, we can hope to probe the properties of the surrounding matter.   Most importantly, we show this echo is detectable in current Earth-based detectors for typical failed galactic supernovae, with an average detection rate of $\sim$2 events per 10\,kT of detector material (whether it be water, argon, or liquid scintillator) for a failed supernovae at a Galactic distance of 10\,kpc.

The neutrino echo is related to the neutrino halo of  \cite{cherry:13,cherry:21}; the neutrino halo is produced via this same mechanism and is responsible for generating a populations of inward going neutrinos throughout the entire CCSN evolution. Similar to \cite{morinaga:20}, where it is shown that coherent scattering in the preshock region leads to an electron lepton number crossing, these works suggest that scattering in the accretion region is of critical importance for the treatment of neutrino oscillations, either collective oscillations or fast-flavor oscillations.  The echo is also related to the work of \cite{nagakura:21}, which was part of the original inspiration for this work.  There, heavy-lepton neutrinos undergo a Fermi-like acceleration as they scatter back and forth across the supernova shock.  This mechanism is no doubt at work in the models we explore here before black hole formation. Furthermore, there is indeed an excess time delay (compared to non-accelerated neutrinos) associated with the increase in the path length of the accelerated neutrinos.  However, such a signal can not extend past the formation of the black hole as the shock is quickly accreted, the acceleration mechanism stops, and the neutrino escape on free streaming trajectories, unless, of course, they then participate in the neutrino echo.

In \S\ref{sec:methods}, we describe the failed core-collapse supernova simulation we utilize in this work as well as introduce the Monte Carlo software we use and modify in order to predict the neutrino echos. In \S\ref{sec:analytic}, we develop an analytic model for the neutrino echo based on coherent scattering in the supersonic accretion flow outside of a protoneutron star.  We present our Monte Carlo neutrino signal results, and compare them to both the original simulations and our analytic predictions in \S\ref{sec:results}. We also present estimates for the Earth-based detector response to the neutrino echos. We assess some potential systematic uncertainties in our neutrino-echo signal predictions due to incomplete model coverage. We also explore, with high-resolution Monte Carlo simulations, the turn off of the neutrino signal immediately following black hole formation. In \S\ref{sec:discuss}, we discuss how the timing profile of our neutrino echo interacts with a potential impact of finite neutrino mass as well as the impact on using the timing of black hole formation for triangulating the location of a Galactic supernovae on the sky.  We summarize in \S\ref{sec:conclude}.

\section{Model and Methods} \label{sec:methods}

\subsection{Model}
\label{sec:GR1D}

\begin{figure*}
\begin{center}
\includegraphics[width=\linewidth]{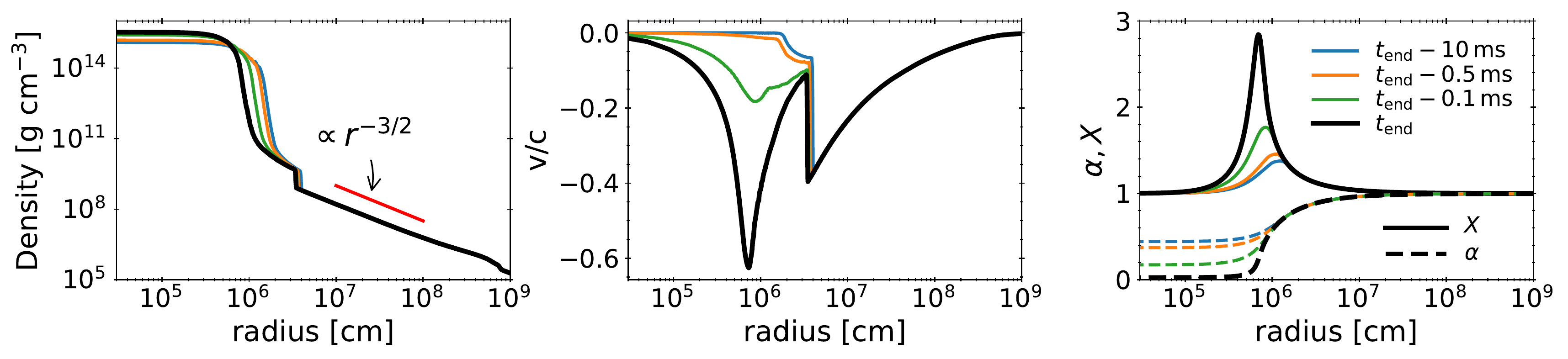}
\caption{Density (left panel), physical velocity ($\sqrt{v_r v^r}/c$; middle panel), and the lapse and radial component of the metric profiles (right panel) at $\sim$10\,ms, $\sim$0.5\,ms, and $\sim$0.1\,ms from the end of the GR1D simulation as well as the last profile from the GR1D simulation.  As GR1D uses a singularity avoiding metric, the black hole never actually forms.  The dynamical collapse of the protoneutron star happens on a very fast time scale, $\lesssim 1$ms and reaches infall velocities of 60\% of the speed of light. Prior to this collapse, the spacetime and hydrodynamic variables only slowly evolve.}
\label{fig:matter_props}
\end{center}
\end{figure*}

We base our Monte Carlo simulations of the evolution of a $40\,M_\odot$ progenitor from \cite{woosley:07}.  This evolution was performed with GR1D \citep{oconnor:10,oconnor:15} and presented in \cite{oconnor:15}.  We refer the reader there for details of the evolution prior to the protoneutron star collapse at the end of the simulation.  The model used the Lattimer and Swesty equation of state \citep{lseos:91} (with an incompressibility modulus of 220\,MeV) and neutrino opacities from NuLib \citep{oconnor:15}.  The evolution was carried towards black hole formation as much as possible, ultimately reaching a post-bounce evolution time of $\sim$537\,ms, but given the form of the metric in GR1D, evolving precisely to the point of horizon formation is impossible, although one can get arbitrarily close within numerical constraints (for example, in our case, due to convergence in the neutrino transport solver).  At the end of the simulation the central value of the lapse has reached $\sim$0.023, the central density has risen to $\sim 3.4\times 10^{15}$\,g\,cm$^{-3}$, and the accretion rate onto the central object is $\sim 0.7\,M_\odot\,\mathrm{s}^{-1}$.  Unfortunately, this is not as far as the evolution presented in \cite{baumgarte:96}, where the simulation was carried out until the lapse throughout the entire protoneutron star was $\sim 10^{-5}$, essentially freezing all subsequent evolution. We quantify the impact of this on the neutrino echo and show it is negligible, but it is nonetheless one limitation of our work. In Figure~\ref{fig:matter_props}, we show profiles of density, the physical velocity, the lapse, and the radial component of the metric at $\sim$10\,ms, $\sim$0.5\,ms, $\sim$0.1\,ms before the end of the simulation.  We also show the very last profile generated in GR1D. The final protoneutron star collapse is quite rapid, with almost all of the dynamics occurring in the final $\lesssim$0.5\,ms. 

For clarity, we describe with some detail how the neutrino luminosity and mean energies are extracted from our GR1D simulation. Following \cite{muller:10} and \cite{oconnor:15}, the luminosity measured at infinity can be related to the conserved flux at any other radius given as,
\begin{equation}
L_\nu = (4 \pi r)^2 \alpha(r)/X(r)^2 \int F^\mathrm{lab}_\nu(r) d\epsilon 
\end{equation}
\noindent
where $\alpha(r)$ is the radius-dependent lapse, $X(r)$ is related to the radius-dependent radial component of the metric, $g_{rr} = X^2$, and $F_\nu^\mathrm{lab}(r)$ is the first moment of the neutrino energy distribution in the coordinate (or laboratory frame).  Note that this expression differs slightly from the luminosity shown in \cite{oconnor:15}, which was just the laboratory luminosity measured at 500\,km. At 500\,km the remaining general relativistic effects, like the red-shift are small, however, in this work we extract the neutrino luminosity at 50\,km, and therefore a proper extraction to the asymptotic value of the luminosity is needed. Likewise, for the neutrino average energy, 
\begin{equation}
\langle \epsilon_\nu \rangle = \alpha(r) W(r) (1+v(r)/c) \langle \epsilon_\nu \rangle^\mathrm{fluid} (r)
\end{equation}
\noindent
where $W(r)$ is the Lorentz factor, $v(r) = X(r) v^r(r)$ is the local fluid velocity, and $\langle \epsilon_\nu \rangle^\mathrm{fluid} (r)$ is the average energy in the fluid frame at radius $r$.  Here we start with the fluid frame quantity since the energy of the neutrinos in each energy bin is defined in the fluid frame.

\subsection{SedonuGR}

SedonuGR \citep{richers:15,richers:17} is an open-source Monte Carlo neutrino transport package.  Given a background spacetime metric, matter profiles, and a set of neutrino interactions, SedonuGR thermally emits and propagates neutrino packets in 3D until they escape the domain.  The neutrinos undergo intermediate interactions along the way including scattering.  Absorption is handled by continuously absorbing some of the neutrino packet energy as the particle moves.  It can utilize 1D, 2D, or 3D matter distributions. The spacetime metric, hydrodynamic variables, and the neutrino-matter interaction coefficients are updated as the particle is transported through the domain.  While the coordinate time of the neutrino is tracked as it is transported, SedonuGR is fundamentally time independent.  It excels at predicting the emitted neutrino spectrum, using exact transport methods, for a steady-state background.

\subsection{Using SedonuGR with time dependent matter profiles}

As shown in Figure~\ref{fig:matter_props}, prior to black hole formation, the spacetime and the hydrodynamic variables in the protoneutron star are only slowly evolving.  Here, the neutrino transport time scale ($\lesssim$ 0.5\,ms to escape the protoneutron star for the majority of neutrinos emitted outside of the core) is short compared to the time over which the hydrodynamic variables are changing, and hence a time independent approximation in SedonuGR is well justified.  However, once the protoneutron star starts to dynamically collapse, the timescales become much shorter, $\lesssim 0.1\,$ms. If one is interested in the neutrino transport at this time, one needs to have a spacetime metric and matter profiles that change dynamically as the neutrino propagates.  We accomplish this with the following relatively modest changes to SedonuGR: 
\begin{enumerate}
    \item When a neutrino is emitted, we first assign a random coordinate start time within a specified time window ($\Delta t$) rather than simply $t=0$. The amount of energy in the neutrino packet is set by the local conditions (in space and time) at the point of emission.  
    \item We promote all of the spacetime, matter, and neutrino interaction fields to be time-dependent.  Whenever a particle is emitted, and for each propagation step, the particle coordinate time is used to select which time profile to use when interpolating the metric, matter fields, or neutrino interaction coefficients. Our time profiles are sufficiently close together to ensure a smooth evolution. 
    \item When a particle escapes, we bin the final distribution in not only neutrino species and momentum space, but also in time with spacing of $\delta t$. 
    \item In the time independent version of SedonuGR, the emission time window ($\Delta t$) and the time observation window ($\delta t$) are taken to be 1 second, although since it is steady-state, the particular values are arbitrary. In our case, the final spectra in each final time bin needs to be normalized according to the $\Delta t$ of the emission and the $\delta t$ of the final timing profile, i.e. $\Delta t/ \delta t$. 
\end{enumerate}

With these changes we are able to model the neutrino emission and determine the energy dependent neutrino signal at Earth not only up to the formation of the black hole (and compare with the M1 results from \citealt{oconnor:15}), but also through and past black hole formation. This is the first detailed neutrino signal prediction in such a scenario since the work of \cite{baumgarte:96} but includes a modern progenitor; a nuclear EOS that can support 2\,$M_{\odot}$; energy dependent transport; and most important for our study, transport that can model neutrino scattering via a Monte Carlo solution of the Boltzmann equation.  It is therefore invaluable for detailed neutrino detection estimates.

\section{Analytic Model for Neutrino Echos}
\label{sec:analytic}
The fundamental interaction occurring to generate the neutrino echos is Freedman scattering of neutrinos on heavy nuclei \citep{freedman:74}.  For the purposes of our analytic model we will take the total coherent neutrino-nucleus scattering cross section to be (ignoring angular dependence and correction terms for the sake of clarity) \citep{burrows:06},

\begin{equation}
    \sigma_{\nu A} =  \frac{\sigma_0}{16}\left(\frac{\epsilon_\nu}{m_e c^2}\right)^2  A^2 \mathcal{W}^2 = \chi_{\nu A} \times  \epsilon_\nu^2 \,,\label{eq:sigmanuA}
\end{equation}

\noindent
where $\sigma_0=1.761\times 10^{-44}\,\mathrm{cm}^2$, $\epsilon_\nu$ is the neutrino energy, $m_e$ is the electron mass, $A$ is the atomic mass of the nucleus the neutrino is scattering off of and $\mathcal{W} = 1 - \frac{2Z}{A}(1-2 \sin^2(\theta_W))$ with $Z$ being the atomic number of the nucleus and $\theta_W$ being the Weinberg angle, taken as $\sin^2(\theta_W)=0.23$. In our model, we take the number spectrum of streaming neutrinos as $dN_\nu(\epsilon)/d\epsilon$. From this we can define the total number and energy luminosity as,
\begin{eqnarray}
N_\nu = \int \frac{dN_\nu(\epsilon)}{d\epsilon} d\epsilon\,,\\
L_\nu = \int \epsilon  \frac{dN_\nu(\epsilon)}{d\epsilon} d\epsilon\,,
\end{eqnarray}
\noindent and the average neutrino energy and higher order moments as,
\begin{eqnarray}
\langle \epsilon_\nu \rangle = L_\nu / N_\nu\,,\\
\langle \epsilon_\nu^n \rangle = \int \epsilon^n \frac{dN_\nu(\epsilon)}{d\epsilon} d\epsilon \Big /N_\nu\,.
\end{eqnarray}

Finally, we model the spherical accretion flow by a power law in mass density of $\rho(r) = \rho_0 (r/r_0)^{-3/2} \exp{(-r/r_1})$, and, for simplicity, with a composition of 100\% iron. The power law index of $-3/2$ is characteristic of free falling material, i.e. compression of a free falling mass element gives a radial profile index of $-2$, but the differential radial stretching of the mass element increases the power law index to $-3/2$ (Also see the left panel of Figure~\ref{fig:matter_props}). Our neutrino interactions turn off at a radius of $r_1=\sim$5000\,km, where the density $\rho$ is $10^6\,\mathrm{g}\,\mathrm{cm}^{-3}$, therefore for the purposes of this model, we include an exponential cutoff. We note however that naturally the density profile turns over at large radii (at the edge of the supersonic region) and becomes steeper than $r\propto -3/2$.   By combining the cross section, neutrino number flux, and target density, and integrating over neutrino energy, we can estimate the scattered luminosity per unit volume,

\begin{equation}
    \frac{d L_\nu^S(r)}{dV}  = \frac{N_\nu}{4 \pi r^2} \langle \epsilon_\nu^3 \rangle \chi_{\nu A} \frac{ \rho(r) N_A}{A}\,,
\end{equation}
where $N_A$ is Avogadro's number and $A$ is the atomic mass, in our case $A=56$ for iron. We note we are ignoring general and special relativistic terms as well as any non-linear feedback on the emitted neutrino spectrum (i.e. we assume only a small fraction of the neutrinos are scattered) for this analytic approximation. Assuming a spherical geometry, we get the scattered luminosity per unit radial length,
\begin{equation}
    \frac{d L_\nu^S(r)}{dr}  = N_\nu \langle \epsilon_\nu^3 \rangle \frac{\chi_{\nu A} \rho_0 N_A }{A}   \left(\frac{r}{r_0}\right)^{-3/2} \exp{\left(-\frac{r}{r_1}\right)}\,.
\end{equation}

If a neutrino is emitted from the origin at $t=0$ and scatters into an observer's line of sight at a radius of $r$, then the travel time to an observer can be delayed by as little as zero time (in the case of forward scattering) or up to $2r/c$, in the case of backward scattering.  For our purposes in this analytic estimate, we will assume if the neutrino scatters at radius $r$, it contributes to the delay time distribution at a time of $r/c$. We assume a minimum scattering time of $\sim r_\mathrm{shock}/c$, which prevents the evaluation of the scattered luminosity at $t=0$, where it is singular.  This gives a delay time distribution for a single burst of neutrino emission at $t=0$ of
\begin{equation}
    \frac{d L^\mathrm{S, single}_\nu(t)}{dt}  = N_\nu \langle \epsilon_\nu^3 \rangle \frac{\chi_{\nu A} \rho_0 N_A}{A} \left(\frac{r_0^3}{ct^3}\right)^{1/2}   \exp{\left(-\frac{ct}{r_1}\right)}\,.\label{eq:tm32}
\end{equation}

However, we do not have just a single burst of neutrinos, but rather a constant source of neutrinos up until the black hole formation time, and hence a convolution of the emission profile and the delay time distribution. In the following, we take $t=0$ to be the black hole formation time and derive the time delay distribution after black hole formation. For simplicity, we assume the spectral properties of the emission are constant, which to the level of this approximation is valid. This convolution gives

\begin{eqnarray}
\nonumber
L_\nu^S(t) &=& \int\limits_{-\infty}^{0} \frac{d L^\mathrm{S, single}_\nu(t-\tau)}{dt} d\tau\,,\\
\nonumber
&=& N_\nu \langle \epsilon_\nu^3 \rangle \frac{\chi_{\nu A} \rho_0 N_A}{A}\left(\frac{r_0^3}{c}\right)^{1/2} \times\\
&&\int\limits_{-\infty}^{0} (t-\tau)^{-3/2} \exp{\left(-\frac{c(t-\tau)}{r_1}\right)}  d\tau\,,
\end{eqnarray}
\noindent
where $\tau$ is the integration variable over the time when the source is emitting. Here we take the lower limit of $\tau=-\infty$ as the exponential decay of the density profile prevents neutrinos emitted before $t \sim -r_1/c \sim -15\,$ms from contributing to the neutrino echo.   This expression evaluates to
\begin{eqnarray}
\nonumber
L_\nu^S(t) &=& N_\nu \langle \epsilon_\nu^3 \rangle \frac{\chi_{\nu A}\rho_0 N_A}{A} \left(\frac{r_0^3}{c}\right)^{1/2} \times\\
&&2\sqrt{\frac{1}{t}}\left [\exp{\left(-\frac{ct}{r_1}\right)} - \sqrt{\frac{ct}{r_1}} \Gamma(\frac{1}{2},\frac{ct}{r_1}) \right
]\,,\label{eq:neutrinoecholum}
\end{eqnarray}
\noindent
where $\Gamma(\frac{1}{2},\frac{ct}{r_1})$ is the upper incomplete gamma function. In the limit of $r_1 \to \infty$, the term in the  $[...]$ is 1 and $L_\nu^S(t) \propto t^{-1/2}$.

As the coherent neutrino-nucleus scattering cross section depends strongly on energy ($\sim \propto \epsilon_\nu^2$), the neutrino echo will have a significantly different spectrum from the emission source. The moments of the energy distribution of the scattered neutrinos will be
\begin{equation}
    \langle \epsilon^n \rangle^{S} = \int \epsilon^{2+n} \frac{dN_\nu(\epsilon)}{d\epsilon} d\epsilon \Big / \int \epsilon^{2} \frac{dN_\nu(\epsilon)}{d\epsilon} d\epsilon\,.\label{eq:neutrinoechoaveE}
\end{equation}
For these moments, since the emission spectra have similar energy moments, and there is no spatial or temporal dependence on the energy, we expect constant energy moments of the scattered neutrinos.  The one exception to this is the special relativistic neutrino-Compton scattering effect, and potentially the neutrino acceleration effect of \cite{nagakura:21}.  The velocity of the accretion flow can be seen in Figure~\ref{fig:matter_props}: interior to a radius of 200\,km, the infall speed is $\sim 0.2\,c$, and the peak value at the shock is $\sim 0.4\,c$. The corresponding Lorentz factors, $\gamma$, are upwards of $\sim$1.1.  Therefore the neutrino-Compton scattering can in principle increase the neutrino energy by factors up to $\gamma^2 \sim 1.2$.  If the optical depth for these neutrinos is high enough, multiple shock crossings could accelerate the neutrinos to even higher energies \citep{nagakura:21}  This would only occur for neutrinos scattered close to the shock (where the velocity is largest) or accelerated before the black hole forms. Therefore we expect any impact of these relativistic effect to be seen with the shortest time delays.

\section{Results}
\label{sec:results}

\subsection{Monte Carlo Results}
\label{sec:main}

\begin{figure*}
\begin{center}
\includegraphics[width=\linewidth]{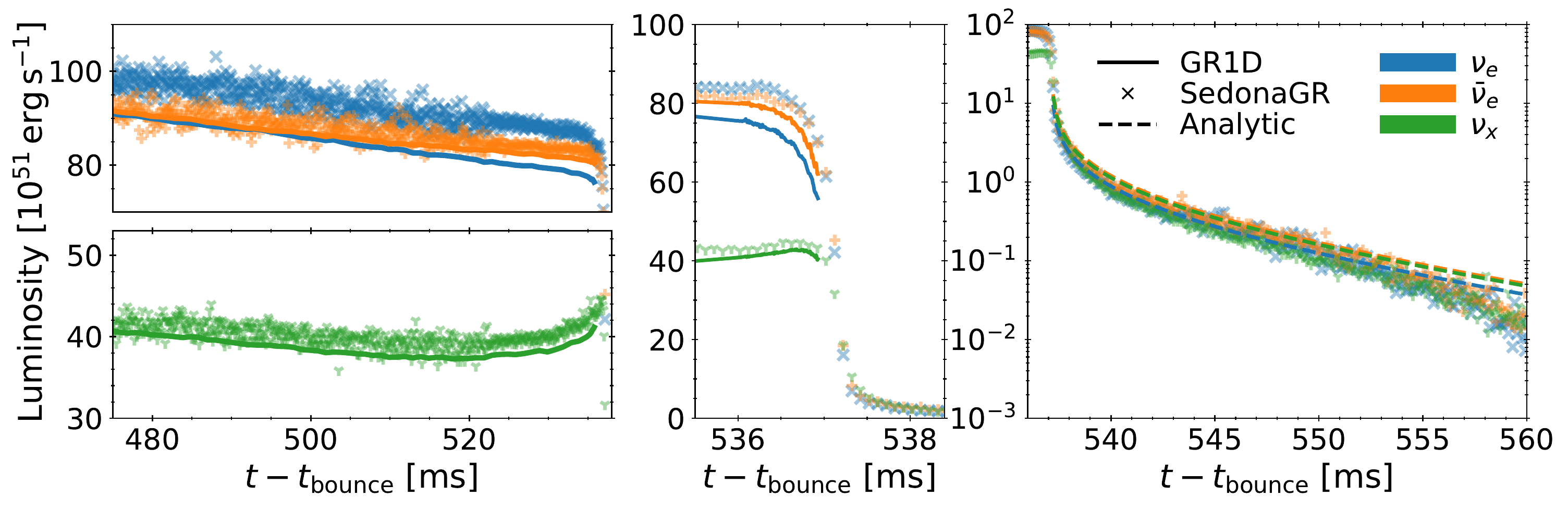}
\includegraphics[width=\linewidth]{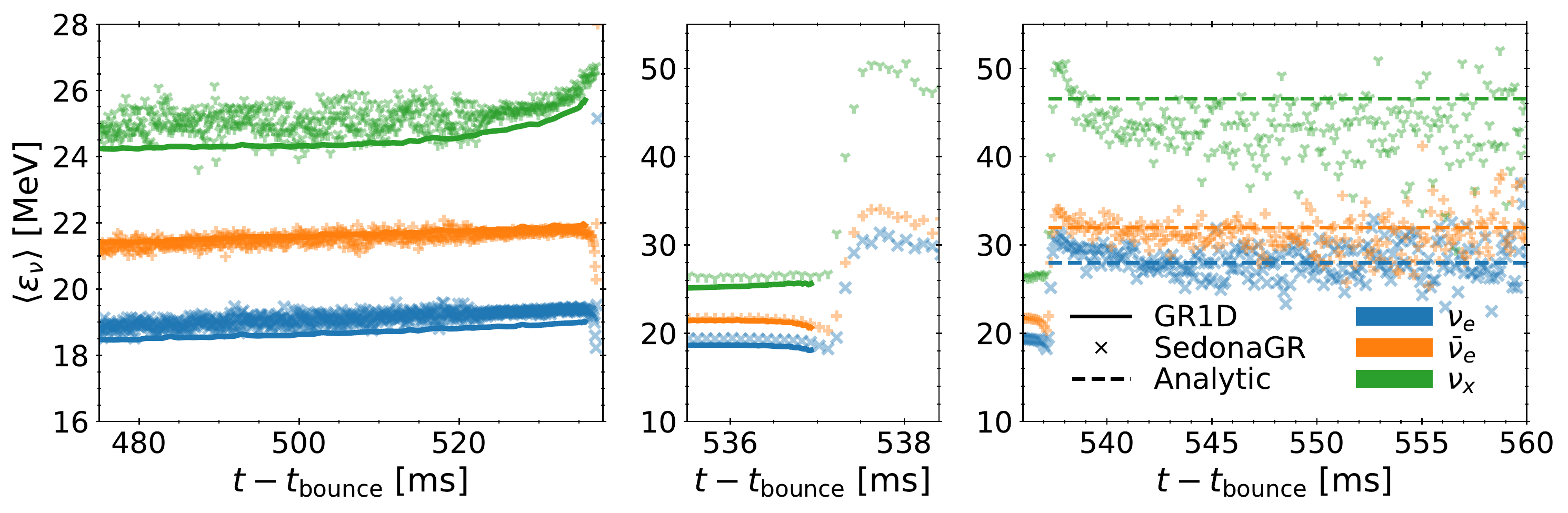}
\caption{Luminosity (top panels) and average energy (bottom panels) of neutrinos before (left panels) during (middle panels) and following (right panels; note the luminosity is in logarithmic scale here) black hole formation.  We show three species, $\nu_e$ (blue), $\bar{\nu}_e$ (orange), and $\nu_x$ (green).  Monte Carlo results are shown with $\times$'s, GR1D results are shown with solid lines (extract at 400\,km in the left panels and 50\,km in the middle panels), and the analytic estimates from \S\ref{sec:analytic} are with dashed lines. Prior to 522\,ms, a lower Monte Carlo particle count is used, which increases the scatter in the Monte Carlo results.  }
\label{fig:s40_withMC}
\end{center}
\end{figure*}

In Figure \ref{fig:s40_withMC}, we show the main results of our study. In the top panel we show the neutrino luminosity as a function of time near, around, and past black hole formation, while in the bottom panel, we show the average energy of the neutrinos. As a note, while we extract the GR1D neutrino signals at particular locations, we show the asymptotic neutrino properties, i.e. at infinity where the lapse and fluid velocity are zero; see \S\ref{sec:GR1D} for details. The Monte Carlo neutrino properties are measured at $R \sim 8\times10^9$\,cm (for all intents and purposes infinity) and time-shifted to line up with the extraction radius $r$ of the GR1D simulation using a time delay of \citep{ames:68},

\begin{equation}
    t_D = \frac{1}{c} \int_r^R X^2 dr\label{eq:timedelay}\,.
\end{equation}

In the left panels, we show the neutrino properties from $\sim$60\,ms before the collapse of the protoneutron star up to the collapse. The solid lines denote the luminosity extracted at 400\,km from the GR1D simulation in \cite{oconnor:15} which used M1 neutrino transport to model the neutrinos. As there is some delay from the neutrino transport, the end of the neutrino signal shown here is carrying information from $\Delta t \sim$ 400\,km/c $\approx$ 1.3\,ms before black hole formation. Since the protoneutron star is still essentially dynamically stable, this signal (and the signal presented in \cite{oconnor:15} that was extracted at 500\,km) shows little sign of the impending collapse. The SedonuGR Monte Carlo results are shown with $\times$ symbols every 0.1\,ms. Prior to $\sim$522\,ms, we simulate 200 emitted Monte Carlo particles per ms, per energy group, per spatial zone, and per species. Following $\sim$522\,ms, we increase this by a factor of 10 to 2000, and in the last $\sim$1.5\,ms, we increase this by a further factor of $\sim$2.5.  In total, we simulate over $10^9$ neutrino particles in SedonuGR. Apart from offsets in the amplitude, the qualitative structure of the SedonuGR signal follows GR1D closely, notably the slowly decreasing amplitude of the electron type neutrinos and the decreasing, then sharply increasing structure of the heavy-lepton neutrinos. We attribute the offsets, which are $\sim$10\% for the electron neutrino luminosity and much less for all the other quantities, to differences in the transport methods, and the neutrino interactions and their angular dependence treatment between the M1 scheme and the Monte Carlo. In particular, we do not include neutrino-electron inelastic scattering in the Monte Carlo simulations as it is not fully implemented in SedonuGR.

In the middle panels of Figure~\ref{fig:s40_withMC}, we zoom in on the time surrounding the protoneutron star collapse to a black hole. For these panels we extract the neutrino signal from the GR1D simulation at 50\,km, allowing us to capture more of the neutrino signal than in the left panels. This was not possible at earlier times since, for example, the shock was located at a larger radius and neutrino emission and absorption was still occurring. At these late times the shock radius is $\sim$35\,km and an extraction at 50\,km is possible, although, since the neutrinos are not fully free streaming, there may be some residual errors in extrapolating the values to infinity. This is not an issue with the Monte Carlo as we simply freeze the spacetime at the point of black hole formation and let the neutrinos transport out to the edge of the domain.  Over the time scale of the 10s of ms explored here, the mass accretion into the newly formed black hole is tiny compared to its total mass, and therefore this is a fine approximation. To handle the matter distribution outside the would-be black hole, we simply place an inner edge of the domain at 10\,km starting at the time of the last GR1D profile. This will allow any neutrino located outside 10\,km to complete its evolution (those inside are instantly absorbed), whether it be eventual escape, or absorption, albeit in a frozen matter field. We assess the impact of this assumption below in \S\ref{sec:impact} where we compare the results to those obtained when removing all neutrinos inside of the shock ($\sim$35\,km). Fully removing the need of this ad hoc treatment would require further hydrodynamical evolution in GR1D, which due to numerical convergence limitations is outside the scope of this work. Nevertheless, in the last $\sim$0.5\,ms of the main neutrino signal we can start to see the effects of the collapsing protoneutron star. As the neutrinospheres begin to shrink and collapse deeper into the gravitational well of the collapsing protoneutron star, both the luminosity and the average energy of the electron-type neutrinos being emitted in these last fractions of a millisecond begin to plummet. The heavy-lepton neutrino luminosity has a slower reaction to the collapse, but eventually turns over as well. It is interesting to note that this behavior is not what is seen in \cite{baumgarte:96}, where instead, the electron neutrinos were the slowest to react to the collapsing protoneutron star. While this could be due to difference in the progenitor in combination with the equation of state, it could also be due to the increased fidelity of the neutrino transport methods used here.

At $t=0.5372$\,ms the average energy of the neutrinos stops decreasing and begins to increase.  We denote this time as when the neutrino echo signal begins to dominate over the decaying protoneutron star signal.  We explore this transition phase in more detail below in \S\ref{sec:atbh}.

In the rightmost panels, we show the main neutrino signal after the formation of the black hole.   Here there is no data from the GR1D simulation, but we show the analytic estimates from \S\ref{sec:analytic} as dashed lines. We take the neutrino fields from the GR1D simulation at 535\,ms to set the number luminosity, $N_\nu$, and the mean energy moments, $\langle \epsilon_\nu^n \rangle$ for each species. The analytic estimates of the neutrino signal, both the luminosity and the average energy, agree remarkably well with the Monte Carlo results. The timing profile is initially shallower than $t^{-3/2}$, which is expected for a neutrino echo from an instantaneous pulse of neutrinos (see Eq.\ref{eq:tm32} for $t\ll r_1/c$), but steeper than $t^{-1/2}$, which is expected from an infinitely extended accretion flow (see Eq.~\ref{eq:neutrinoecholum} with $r_1=\infty$).  The Monte Carlo signal  follows Eq.~\ref{eq:neutrinoecholum} (with $r_1=5000\,\mathrm{km}$) remarkable well. The energy prediction, via Eq.~\ref{eq:neutrinoechoaveE}, also matches the Monte Carlo results well.  The dramatic increase of $\sim$10\,MeV for $\nu_e$ and $\sim$25\,MeV for $\nu_x$, is due not to particular neutrinos gaining energy, but rather the high energy neutrinos preferentially scattering in the accretion flow due to the strong energy dependence of the coherent scattering cross section. These estimates confirm that neutrino echoes, from coherent neutrino-nucleus scattering in the accretion flow, can be a dominant source of neutrinos post black hole formation. We assess our ability to detect these neutrino echos in current and near-future neutrino detectors below in \S\ref{sec:detect}.

There remain some differences between the analytic estimates and the Monte Carlo signal that deserves some discussion. The time dependence of the luminosity of the scattered neutrinos deviates from the Monte Carlo results at later times. This is due to the sharp cutoff in scattering in the Monte Carlo at $\rho=10^6$\,g\,cm$^{-3}$ at $r\sim5000$\,km whereas the model only has an exponential decay. Overall, the amplitude deviation is strongest for the $\nu_x$ neutrinos, which happen to be the most energetic, therefore we suspect that our neglect of the nuclear form factor in the coherent neutrino nucleus scattering cross section within our analytic estimate (it is included in the NuLib opacities used in SedonuGR) may be the cause.  The form factor reduces the scattering cross section for the highest neutrino energies; for iron at 50\,MeV, the cross section is reduced by $\sim$50\% \citep{burrows:06}. This could be incorporated into our analytic model with the added complexity of a cross section that does not simply scale with $\epsilon_\nu^2$. The nuclear form factor could also explain the slightly higher predicted average energies (again, largest for the $\nu_x$).  Another difference we note is the slightly larger neutrino energy just following the black hole formation.  This is seen for all species.  As alluded to in \S~\ref{sec:analytic}, the neutrinos that arrive very soon after the black hole formation likely have a shorter time delay, and therefore under went the scattering at a lower radius.  Here the matter velocity is mildly relativistic and can Doppler boost the energy of the neutrino. With the maximum fluid velocity of $0.4\,c$ (see Figure~\ref{fig:matter_props}), this is likely to only increase the energy by at most $\sim \gamma^2-1 \sim 20\%$, roughly what we see here. It is not entirely unlikely that there could be some non-thermally accelerated neutrinos \citep{nagakura:21} present at these early times after black hole formation. Finally, for each of the neutrino species, the Monte Carlo average energy, while quite noisy, is actually increasing at the latest times.  This is due to multiple scatterings in the accretion flow.  Multiple scatterings are increasingly likely for higher energy neutrinos and lead to long delay times.  The amount of these double scatterings is very low as the luminosity is already reduced by a factor of $\sim 10^4$ after $\sim$20\,ms.

\subsection{Detection Estimates for Neutrino Echos from failed CCSNe}
\label{sec:detect}

\begin{figure}
\begin{center}
\includegraphics[width=\columnwidth]{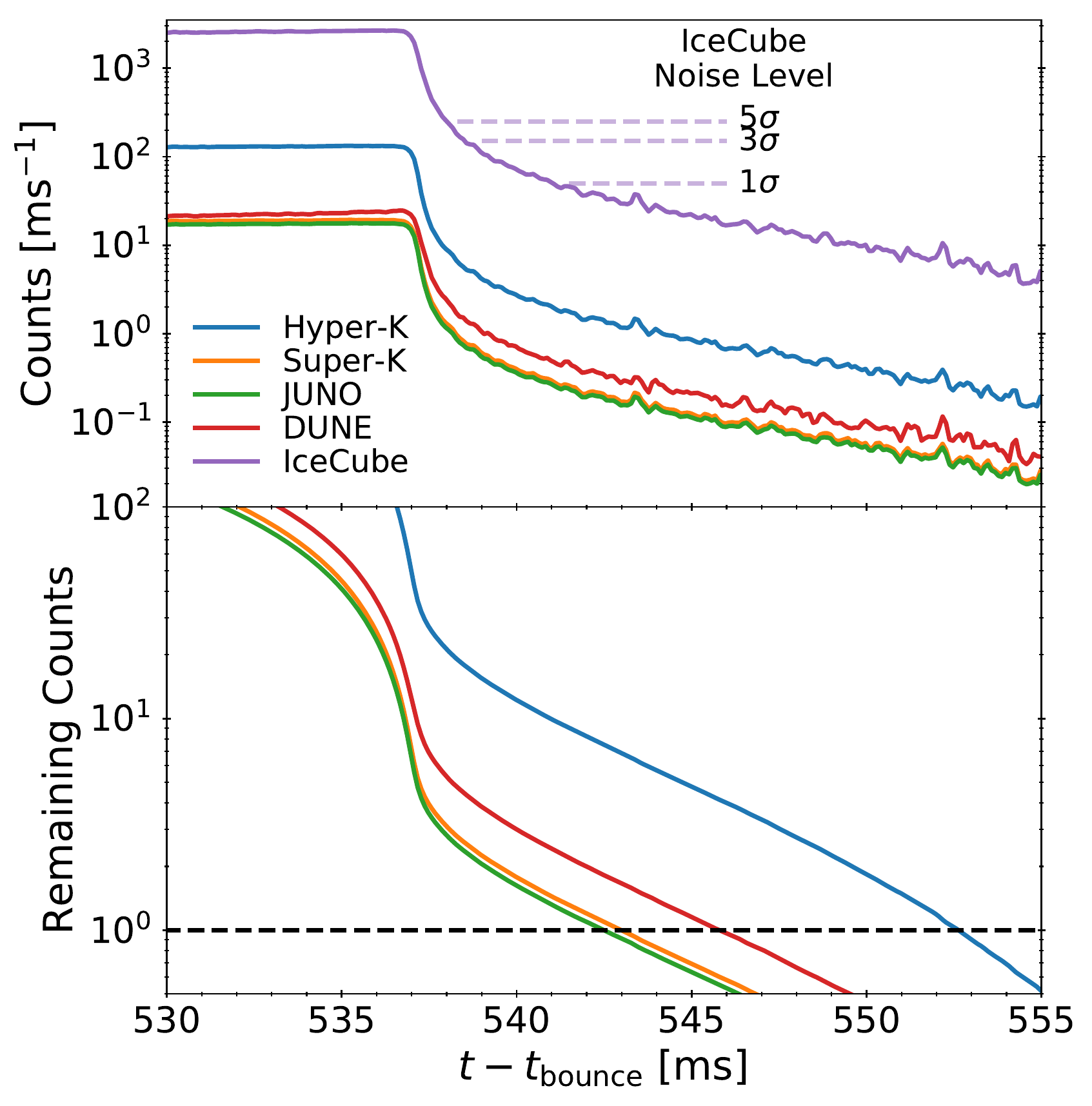}
\caption{Estimated count rate (upper panel) and counts remaining to be detected (bottom panel) in several current and future neutrino detectors for the neutrino echo following the black hole formation in the models explored here.  We assume a distance of 10\,kpc and adiabatic neutrino oscillations assuming the normal mass ordering.  In the top panel we include the 1, 3, and 5$\sigma$ values of the expected background noise level for IceCube, see the text for details.}
\label{fig:snowglobes}
\end{center}
\end{figure}

We now proceed to determine if the neutrino echo signal following black hole formation in a Galactic supernova is detectable in Earth-based detectors.  For this, we extract from the Monte Carlo data, in 0.1\,ms time bins, the luminosity, average energy, and root mean squared energy for each neutrino species in a 30\,ms time window around black hole formation, starting $\sim$7\,ms before and continuing $\sim$23\,ms after black hole formation. With these quantities, we use SNEWPY \citep{snewpy} and SNOwGLoBES \citep{snowglobes} to estimate the interaction rate of our model for two currently operating detectors, a Super-K-like, 32\,kT water Cherenkov detector \citep{superK:2003} and an IceCube-like, long-string water Cherenkov detector \citep{icecube:supernovae:2011}.  We also include three future detectors: a JUNO-like, 20\,kT liquid scintillator detector \citep{juno:physics:16}; a Hyper-K-like, 220\,kT water Cherenkov detector \citep{hyperk:dr:2018}; and a DUNE-like, 40\,kT liquid argon detector \citep{dune:tdr:physics,dune:supernova}.  We assume a Galactic distance of 10\,kpc, and adiabatic MSW oscillations with the normal mass ordering. In Figure~\ref{fig:snowglobes}, we show the interaction rate (top panel) and the dwindling number of neutrinos remaining to be detected (bottom panel; excluding IceCube, see below) for each detector. We summarize the estimated total number of neutrino-echo neutrinos interactions in Table~\ref{tab:number}.  We do not explicitly include the inverted mass ordering results, however we note that the spectral properties of the $\nu_e$, $\bar{\nu}_e$, and $\nu_x$ neutrino are coincidentally such that both mass orderings give similar results (within a few percent) for all detectors with the exception of IceCube which gives a $\sim10\%$ higher yield when assuming the inverted mass ordering due to the stronger energy dependence of the detector. We note that SNOwGLoBES only extends to 100\,MeV, therefore these rates and count predictions may be an underestimate. For Super-K, JUNO, and DUNE, regardless of the neutrino mass ordering, between 5 and 10 neutrino-echo neutrinos are expected following black hole formation.  We expect these neutrinos to arrive within $\sim$5-7\,ms of the cessation of the main neutrino signal. Hyper-K, with its significantly larger target mass, is expected to see $\sim$35 neutrinos, with the last neutrino arriving after $\sim$15\,ms.  With this large number of predicted neutrinos, properties of the neutrino echo may be able to be constrained from a detection with Hyper-K.  

\begin{table}[ht]
\centering
\begin{tabular}{c|c|c|c|c}
\hline
Detector & Target & Mass & \multicolumn{2}{c}{Counts in Neutrino Echo} \\
&& [kT] & baseline & truncated \\

\hline
\hline
Super-K & Water & 32 & 5.2 & 4.4\\
JUNO & L. Scint. & 20 & 4.7 & 4.0\\
DUNE & Argon & 40 &  9.4 & 7.6\\
Hyper-K & Water & 220 & 35.9 & 30.4 \\
IceCube & Water & 2500* & 950 &800\\
\hline
\end{tabular}
    \caption{Estimate of the number of neutrino-echo neutrino interactions in detectors similar to Super-K, JUNO, DUNE, Hyper-K, and IceCube, based on event rates determined by SNOwGLoBES and the Monte Carlo results for a failed supernovae based on the model explored here at 10\,kpc assuming only adiabatic MSW neutrino oscillations with the normal mass ordering. We include our baseline model presented throughout \S\ref{sec:results}, and also the truncated model discussed in \S\ref{sec:impact}.  We define the start of the neutrino echo to be when the average energy of the neutrinos begin to increase after black hole formation, (see bottom panel of  Figure~\ref{fig:s40_withMC}, $t=0.5372$\,s).  The $*$ on the mass of IceCube refers to the effective mass, we take the value from \cite{snews2p0}. While the effective mass is $\sim$10 times that of Hyper-K, there are $\sim$25 times the events detected from the neutrino-echo, this is because the strong energy dependence of IceCube (compared to Hyper-K) and the very high energies of the neutrino-echo neutrinos. }
    \label{tab:number}

\end{table}

The IceCube signal is unique.  While there are many expected interactions in IceCube, the large background rate prevents a clean detection \citep{icecube:supernovae:2011}.  To quantify the impact of the noise background, we show in the top panel of Figure~\ref{fig:snowglobes} the expected $1\sigma$, $3\sigma$, and $5\sigma$ deviations from the background noise. For reference $1\sigma \sim 50$ counts per ms and is taken as $1.3 \times \sqrt{286\,\mathrm{Hz} \times 1\,\mathrm{ms} \times 5160}$, where 286\,Hz is the background rate per digital optical module (DOM) \citep{lutz:2011} and 5160 is the number of DOMs in IceCube. The number in the square root is the total number of background events in IceCube over a 1\,ms time window, and the square root is the $1\sigma$ deviations based on Poisson statistics.  The factor of 1.3 is taken from \cite{icecube:supernovae:2011} and is used to compensate for the impact of correlated pulses in the detector, although this number is determined for 0.5\,s duration windows, and not the 1\,ms windows as used here.  \cite{icecube:supernovae:2011} suggest that post-processing the data can reduce this factor, although the precise level of the noise is not important for this work.  Within the first 1, 2, and 4 milliseconds after the black hole formation, the rate shown here is $\sim 5\sigma$, $\sim 3\sigma$, and $\sim 1\sigma$ above the background.  This suggests the presence of a neutrino echo may be discernible in IceCube, but characterizing its evolution would be difficult unless the supernova was nearby. 

Any detection of the neutrino-echo neutrinos would be invaluable for characterizing the properties of the matter surrounding the protoneutron star as it collapses.  In particular, based on the analytic expression for the neutrino echo luminosity derived in \S\ref{sec:analytic}, i.e. Eq.~\ref{eq:neutrinoecholum}, a detection could constrain the magnitude of the density profile via $\rho_0$ in combination with the composition. The density is related to the accretion rate, which can vary by up to an order of magnitude at black hole formation, depending on the progenitor \citep{schneider:20}. While we have assumed iron in our analytic approximation, and the composition of the nuclear EOS table we use is also close to iron group nuclei for the densities present in the supersonic infall region, the actual composition of the accretion flow may not be iron, and in fact may be lighter nuclei.  The total coherent neutrino-nucleus scattering opacity is, to zeroth order, proportional to $N^2/A$, therefore replacing the iron with silicon or oxygen would result in a reduced opacity by a factor of $\sim$2 and $\sim$3, respectively. In reality, the composition could be a combination of elements and in general is progenitor dependent.  The time dependence of the signal directly reflects the radial dependence of the density, and therefore a particularly precise measurement would confirm or rule out the presence of a supersonic accretion flow.  This is potentially important because if black hole formation occurs following the successful revival of the supernova shock, the surrounding matter may not have the assumed $\rho(r) \propto r^{-3/2}$ profile. 

\subsection{Impact of artificial cutoff}
\label{sec:impact}

Past the end of the simulated spacetime we do not know the future evolution of the matter and metric.  This means that we cannot consistently evolve the neutrinos forward in time. However, this only impacts a small number of the neutrinos involved in the neutrino echo as the matter profiles and the spacetime outside of the supernova shock, the supersonic infall regime, do not significantly evolve over the timescales of interest here.  Therefore it is only the matter, and the neutrinos that traverse it, in the protoneutron star and the post-shock region that are of concern.  We make two assumptions to assess the impact of this uncertainty on the neutrino echo signal.  Our default choice, and the one used above, is to freeze the matter and metric profiles to the values obtained at the end of the GR1D simulation and to place an inner boundary at 10\,km. Any neutrino within the boundary, or interacting with the boundary is absorbed. This allows neutrinos that are outside 10\,km to still propagate and contribute to the signal at Earth. The density at 10\,km is $\sim 10^{11}$\,g\,cm$^{-3}$ (see Figure~\ref{fig:matter_props}), and therefore many neutrinos outside of this 10\,km core will not have any interactions with the matter and simply free stream away. For example, the scattering optical depth from 10\,km, 20\,km, and 35\,km to infinity for a 30\,MeV neutrino is 1, 0.2, and 0.07, respectively.  A conservative assumption, and one that we make to overestimate the overall impact of the uncertainties associated with the final matter profiles, is to place the inner boundary outside of the supernova shock (at 35\,km), this completely removes the hydrostatic structures (and any lingering neutrinos) inside the shock and leaves the supersonic accretion flow outside.  The downside of this is we lose the last $\sim 25\,\mathrm{km}/c \sim 0.1\,\mathrm{ms}$ of emitted neutrinos, which carry critical information regarding the collapsing protoneutron star. For times where the neutrino echo signal is unaffected by these assumptions, we are confident in our ability to model it free of these model uncertainties.

\begin{figure}
\begin{center}
\includegraphics[width=\columnwidth]{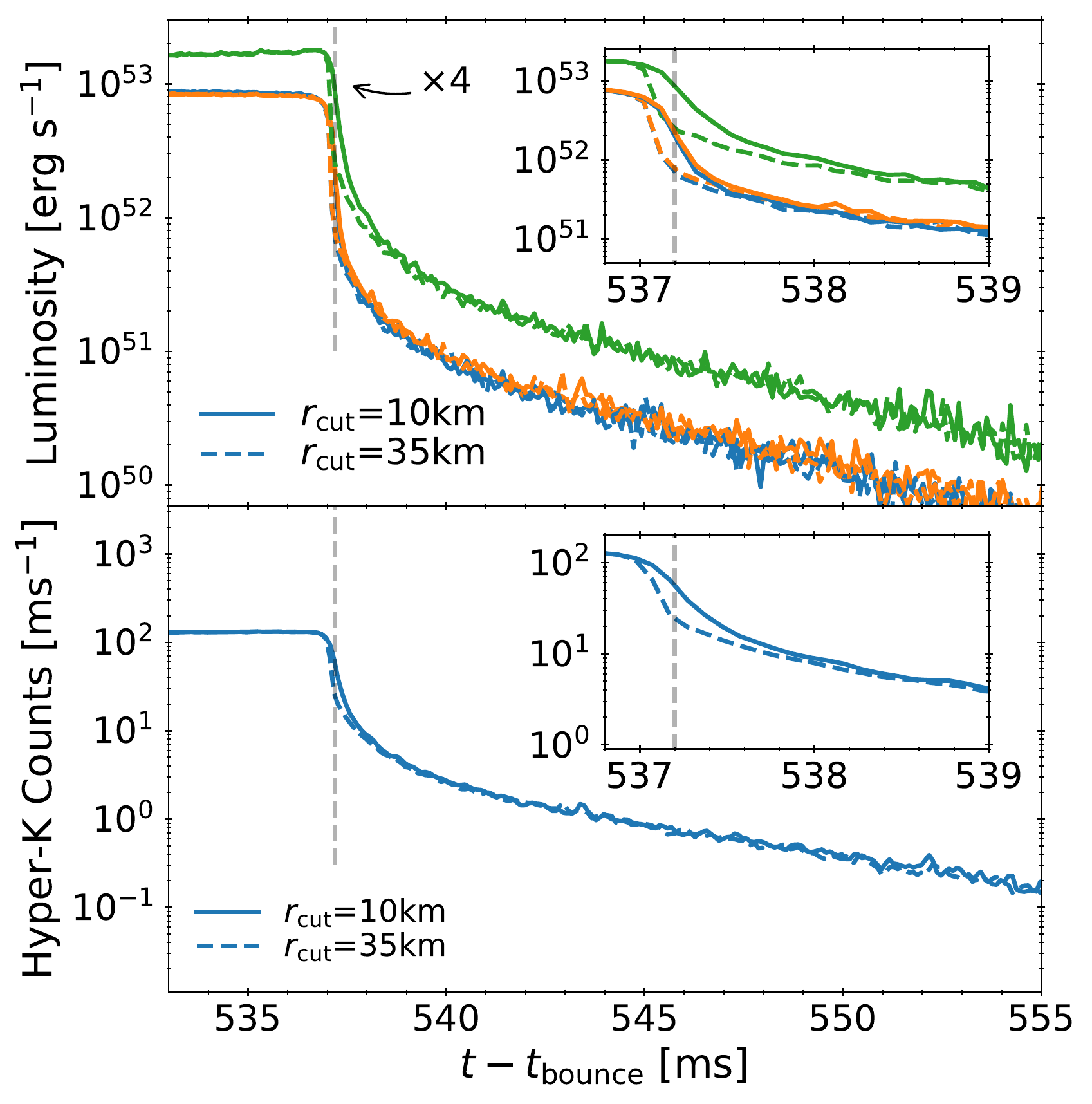}
\caption{Impact of the cutoff models used to continue the matter and spacetime profiles past the last coordinate time of the GR1D simulation. The top panel shows the neutrino luminosity for the three neutrino species and the bottom show the estimated interaction rate in a Hyper-K-like detector.  The solid lines are our fiducial model and the dashed lines correspond to a truncated model where we remove any neutrino located with 35\,km when its coordinate time surpasses the time of the last profile from GR1D. The truncated model removes any uncertainty arising from this unknown evolution at the expense of removing usable neutrinos. For clarity, in the top panel and its insert we show the full $\nu_x$ luminosity, which is four times the single-species value shown in Figure~\ref{fig:s40_withMC}. The vertical dashed line is at a time of $t=0.5372$\,s, the time at which the average energy of the neutrinos begins increasing and the neutrino-echo begins to dominate the signal.}
\label{fig:impact}
\end{center}
\end{figure}

In Figure~\ref{fig:impact}, we show the neutrino luminosity (top panel) and estimated Hyper-K interaction rate (bottom panel) during and after the black hole formation (the same as used in \S\ref{sec:detect} for assessing the detectability of the neutrino signal). We include our fiducial model (solid lines) and the truncated model (dashed lines) where we take the drastic step to remove any neutrino inside the supernova shock at 35\,km when the coordinate time reaches our last profile. The main difference is the extended emission in the fiducial model for $\sim$0.2-0.4\,ms past the sharp cutoff seen in the truncated model.  For each panel in Figure~\ref{fig:impact} we show an inset focused on this time. This extended emission mainly corresponds to the free streaming neutrinos located between 10\,km and 35\,km that we note are unlikely impacted by the model uncertainties and legitimately contribute to the signal at Earth.  The heavy-lepton neutrinos are more extended than the electron type neutrinos.  We expect this is due to high energy heavy-lepton neutrinos that scatter in the protoneutron star and post-shock region before escaping.  The correspondingly high energy electron-type neutrinos (of which there are fewer due to the lower average energy) are absorbed via charged-current interactions, which are absent for the heavy-lepton type neutrinos. The main result here is that by $\sim$1\,ms after the black hole formation the neutrino echo signal is identical in both models.  The majority of the predicted neutrino-echo neutrinos arrive after this 1\,ms.  For a quantitative assessment, we include estimates for the number of detected neutrino-echo neutrinos from the truncated model in Table~\ref{tab:number}.  The number is necessarily smaller than the fiducial model.  A fraction of the difference will be neutrino-echo neutrinos emitted very near black hole formation and that scattered at very small radii, the remaining neutrinos will be late-time decay neutrinos leaking from the vicinity of the photon ring of the black hole.  We explore this neutrino signal at the point of black hole formation in the following section.

\subsection{Neutrinos at black hole formation}
\label{sec:atbh}
\begin{figure*}
\begin{center}
\includegraphics[width=\textwidth]{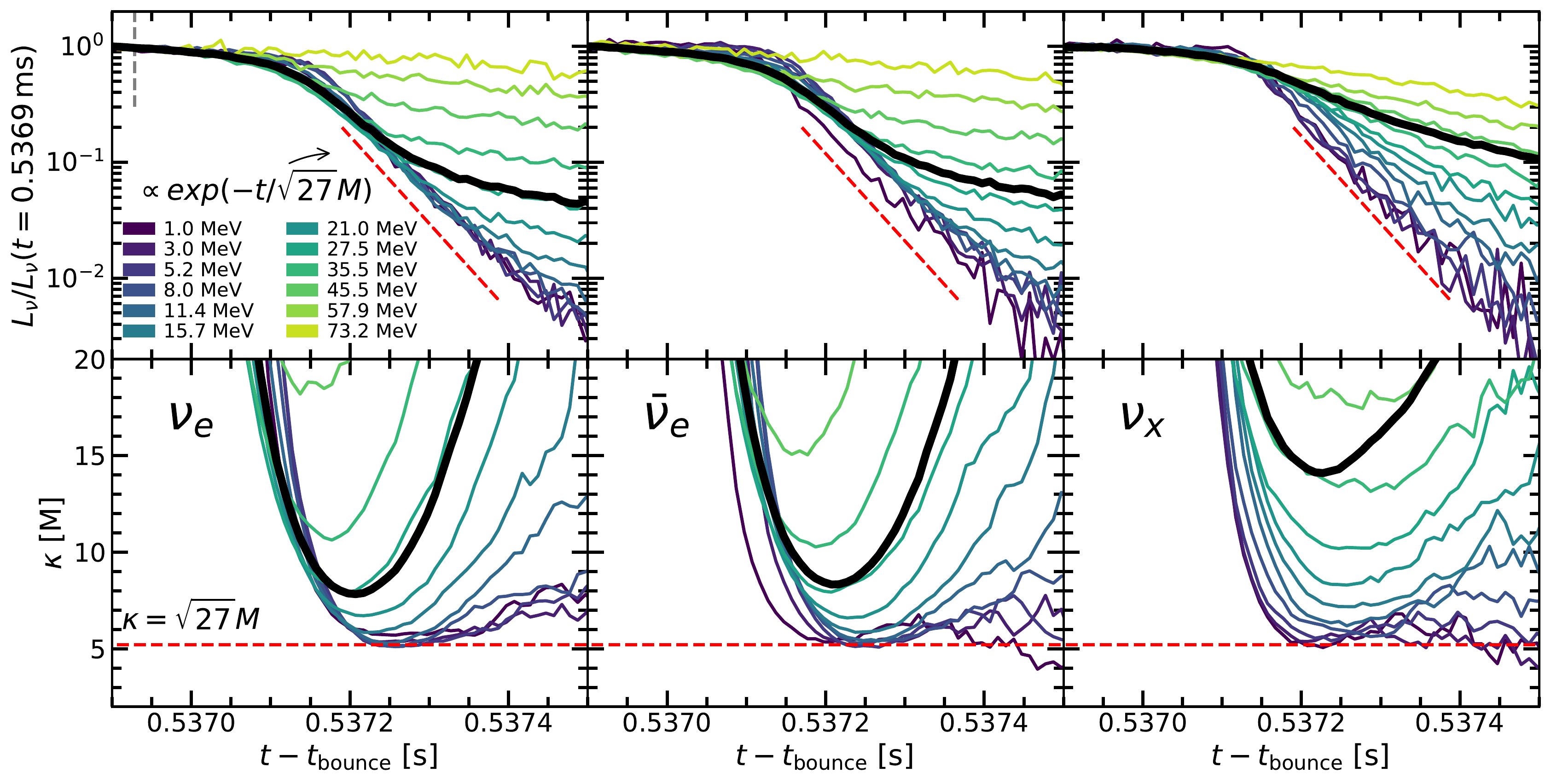}
\caption{Normalized neutrino spectra for individual neutrino energies (top panel) and the associated decay constant (bottom panel) at black hole formation within our fiducial model. We show this for the 12 lowest energy groups for the electron neutrino (left), electron antineutrino (middle), and the characteristic heavy-lepton neutrino (right).  The spectra are normalized to the value at $t=0.5369$\,s, the value at the left side of the figure. In black we show the total luminosity summed over all energy bins.  The decay constant relays the exponential decay constant of the luminosity, i.e. $L_\nu \propto \exp{(-t/\kappa)}$.  We show the expected asymptotic decay of $\sqrt{27}M$ in both panels. While the lowest energy neutrinos achieve this limit, the spectrum of the high energy neutrinos is soon dominated by the neutrino echo and the signal from the decay surrounding the black hole is overwhelmed. The dashed line at $t\sim 0.53693$\,s corresponds to the time of the last matter and spacetime profiles from GR1D when we excise the inner 10\,km of the domain and stop emission.  For consistency with our earlier results, we time shift the Monte Carlo results such that a radial, free-streaming neutrino emitted from 50\,km at the time of the last GR1D profile would register the time of the grey dashed line near the top left corner, see the text for details.}
\label{fig:spec_at_bh}
\end{center}
\end{figure*}

To explore the protoneutron star collapse in detail we use high-time-resolution ($\delta t$ = 0.01\,ms) Monte Carlo simulations at the turn off of the main neutrino signal at black hole formation.  We emit neutrinos starting from $\sim$1.5\,ms before black hole formation and follow their evolution.  Like our fiducial model, when the last matter profile is reached, we remove all neutrinos within 10\,km (corresponds to $\sim$3M), and allow the remaining neutrinos to propagate in the frozen matter field. For consistency with \S\ref{sec:main}, we time shift the Monte Carlo results such that a radial, free-streaming neutrino emitted from 50\,km at the time of the last GR1D profile would register the time of the dashed line shown in Figure~\ref{fig:spec_at_bh}. However, since we allow all neutrinos outside 10\,km to continue to propagate, a neutrino emitted on a radial trajectory from the protoneutron star surface ($\rho=10^{11}$\,g\,cm$^{-3}$) at $\sim$10\,km would arrive $\sim0.2$\,ms later, following Eq.~\ref{eq:timedelay}. For this reason, the strongest decay of the neutrino signal does not begin until $\sim$0.5371\,ms.

In the top panels of Figure~\ref{fig:spec_at_bh}, we show the detailed decay of the total normalized neutrino luminosity in this regime (solid black line) for $\nu_e$, $\bar{\nu}_e$, and $\nu_x$ in the left, middle and right panels, respectively.  In color, starting with the lowest energy groups in blue and increase towards yellow, we show the normalized neutrino luminosity within the first 12 neutrino energy group (at infinity, i.e. the neutrinos are redshifted already).  Eventually, the neutrino echo overwhelms the decaying signal from the collapsing protoneutron star, although this occurs much sooner for the higher energy neutrinos and quite late for the lowest energy neutrinos. The black curve follows closely to the curve corresponding to the average energy of the neutrinos, for example $\sim$30\,MeV for $\nu_e$ and $\sim$45\,MeV for $\nu_x$ in the neutrino echo phase. Note that since we only simulate from $\sim$1.5\,ms before black hole formation, the neutrino echo seen here is incomplete compared to that shown in \S\ref{sec:main}; including more prior evolution would serve to slightly increase the neutrino echo signal (compared to what is seen here), but would be very computationally expensive, given the timing resolution and the resolution for each neutrino energy group we desire for this exploration. 

The decay of the luminosity in Figure~\ref{fig:spec_at_bh} is quantified in the bottom panel. We show the decay constant, $\kappa = -1/(d\ln(L_\nu)/dt)$, in units of $GM/c^3=G \times 2.25\,M_\odot/c^3 \sim 11.08\,\mu s$ in the bottom panel of Figure~\ref{fig:spec_at_bh}.  Like the top panel, the solid black line corresponds to the entire neutrino signal, each color is an individual neutrino energy group.  The decay of the total signal never reaches the asymptotic value of $\sqrt{27} M$ \citep{podurets:64,ames:68, wang:21} owing to the rise of the neutrino echo.  However, we confirm that escaping low energy neutrinos, which are much less likely to participate in the neutrino echo due to the energy dependence of the coherent scattering cross section do indeed reach this limit.  For a given energy group, the heavy-lepton neutrinos do not reach as small values of $\kappa$ as the electron-type neutrinos. We attribute this to the heavy-lepton neutrinos being generally emitted deeper in the star and therefore at a higher optical depth where they will undergo more scatterings and draw out the decay signal.  We note however that this is likely subject to uncertainties in our treatment of the matter after the last matter profile from GR1D, which we discussed in \S\ref{sec:impact}.

We would like to briefly and only roughly address how well one can measure the decay constant $\kappa$. This is of particular interest as a direct measurement of a constant-in-time decay constant would allow for the association with expected analytic value of $\sqrt{27}M$ and therefore a direct measurement of the birth mass of the black hole. Using the detection estimate in \S\ref{sec:detect} for a failed supernova at 10\,kpc, we estimate a detector like Hyper-K will see, from this particular model, $\sim$10 neutrinos between the times of 0.5371\,s and 0.5373\,s, a 200\,$\mu$s window. This window is when the decay constant $\kappa$ is the smallest and therefore when the signal drops the fastest. This is just enough that one may entertain the possibility of measuring the decay constant assuming an exponential decay.  It is important to note, however, that some fraction of these $\sim$10 neutrinos are likely neutrino-echo neutrinos and therefore not involved in the leakage from the neutrino cloud near $r=3M$. Therefore, disentangling these effects and establishing the exponential form of the decay by itself would require more data and likely a simultaneous fit of both decay and echo parameters to the event times and energies near the end of the neutrino output.

\section{Discussion}
\label{sec:discuss}

\emph{Neutrino Masses:} In addition to the dynamical black hole formation causing the neutrino signal to have a turn-off timescale of $\sim0.1\,$ms, finite neutrino mass, which limits the time-of-flight of the neutrinos to less than the speed of light and introduces an energy dependence, can smooth the cutoff also.  \cite{beacom:01} show the time delay compared to massless neutrinos is,
\begin{equation}
    \Delta t = 0.515\, \mathrm{s}\,\left(\frac{m}{E}\right)^2 \frac{D}{10\,\mathrm{kpc}}\,,
\end{equation}
\noindent
where $m$ is the neutrino mass in eV, $E$ is the neutrino energy in MeV, and $D$ is the distance to the supernova. For neutrino masses of 1\,eV, the upper limit on the electron neutrino mass from the direct mass measurement \citep{katrin:19}, and neutrino energies of 30\,MeV, this equates to $\sim$0.5\,ms for a supernova at 10\,kpc. This is on the timescales that would impact the timing structure of the neutrino cutoff. However, with cosmological measurements limiting the neutrino masses to $\lesssim 0.12\,$eV \citep{planck:18cp} which corresponds to time delays for 30\,MeV (5\,MeV) neutrinos at 10\,kpc of only $\sim$0.008\,ms ($\sim$0.3\,ms), actual finite-mass induced time delays into the neutrino echo itself (especially at the energies of interest) are likely negligible.  On the other hand, limits on the neutrino mass from the timing structure of the neutrinos \citep[cf.][]{hansen:20} following a black hole must consider the potential presence of neutrino echos as a background since detectors that can well resolve the shutoff of the main signal, like Hyper-K, may have significant echos.  In the future, for large detectors capable of measuring the decay of neutrinos directly following black hole formation (see \S\ref{sec:atbh}), neutrino mass effects may be important to consider.

\emph{Triangulation:} The initial rise of the neutrino signal at core bounce, or the abrupt cutoff of the neutrino signal at black hole formation can be used to help locate the supernova on the sky via triangulation in multiple detectors \citep{beacom:99,muhlbeier:13,linzer:19,hansen:20}. This relies on different detectors across the globe precisely determining the time of key events, for example the first event \citep{linzer:19}, or the last event in the case of failed supernovae \citep{hansen:20}, and using the difference in the time stamps to place the supernova in a location on the sky.  In addition to statistical concerns comparing detectors of different sizes \citep{beacom:99}, our work here shows that in the case of failed supernovae care must be taken on isolating the timing of the main neutrino signal cutoff. Simply using the last event is insufficient, since the neutrino-echo neutrinos can last for up to 15\,ms following the collapse.  Nevertheless, in detectors with significant neutrino-echo neutrinos, the turn off of the neutrino signal at black hole formation should be well resolved and therefore a precise time of black hole formation should be determinable.

\emph{Multidimensional Effects:} Multidimensional simulations of black hole forming core collapse events can deviate from the spherically symmetric picture shown here.  It is worth discussing these differences and how they may impact the neutrino-echo.  Since multidimensional instabilities dominate the dynamics of the gain region, it can be that the supernova shock is not located at the small radius predicted by our spherically symmetric simulations.  There may be an additional extended neutrino signal from the hot, shocked matter as it accretes into the black hole on a longer time scale, however, as long as there is still a supersonic accretion flow and a stalled shock at a radius $\lesssim 100\,\mathrm{km}$, the neutrino echo is unlikely to be dramatically altered as the bulk of the echo comes from scatterings at larger radii with longer time delays. As mentioned earlier, it can even be that the explosion mechanism was successful and the supernova shock is located at a large radius at the time of black hole formation \citep{chan:18,obergaulinger:17}.  In these scenarios it can be that some of the shock material can even escape to infinity. For these supernovae, the neutrino echo may be significantly altered. The hot, shocked material may reach radii of 1000s of km, altering the matter profiles from what we assumed here (i.e. not $\rho \propto r^{-3/2}$). In such cases, any observed neutrino echo could be used to constrain the structure of the matter fields outside of the newly formed black hole.

\emph{Rotation:} For protoneutron stars that are rotating when they undergo collapse to a black hole, frame dragging of neutrino paths around the black hole can further delay their escape from near the core \citep{wang:21}.  In extreme cases, with angular momentum $J=M^2$ (or more conventionally $a\equiv J/M=M$), the decay in the luminosity departs significantly from the asymptotic value of $\sqrt{27}M$ expected for non-rotating black holes, adding 0.1 to 0.15\,ms to the decay time, which would then feed into the time structure of the neutrino echo.  Of course, such extreme rotation would also modify the dynamics of accretion, and thus the echo itself. Furthermore, for extreme rotation, the formation of a disk that would be presumably a strong source of neutrinos itself may occur.  Observation of significant angular momentum so early in the life of a remnant black hole would have implications for the progenitor, since in the conventional picture of single-star core collapse, the core cannot obtain such high angular momentum \citep{woosley:06}.  At the same time, if it can, it would be expected to undergo a strong magneto-rotational explosion, resulting in a protomagnetar \citep{dessart:12}, which may yet collapse into a rapidly spinning black hole if it continues to accrete matter \citep{aloy:21}.

\section{Conclusion}
\label{sec:conclude}

We present here a new component of the neutrino signal from a failed core-collapse supernova. Following the formation of the black hole the source of neutrino emission is lost, however, neutrino scattering in the medium surrounding the black hole can cause a neutrino echo to form, extending the neutrino signal past the time of black hole formation with a tail lasting up to $\sim$15\,ms.  It is worth noting that this echo is always present, but it takes the abrupt termination of the overwhelming neutrino signal from the protoneutron star for it to dominate. 

We develop a simplified analytic model to predict the spectral properties of the neutrino echo (i.e. the luminosity and mean energy) and the time dependence.  Furthermore, we confirm its presence, and the validity of our model, using Monte Carlo simulations of neutrino transport.  The latter required extending the SedonuGR Monte Carlo code to be semi-time-dependent by taking a time-dependent background matter field and spacetime for the emitted neutrinos to propagate in.  

The neutrino-echo neutrinos are significantly higher in energy than the main neutrino signal neutrinos.  $\nu_e$ have average energies in the echo of $\sim$\,30\,MeV, compared to $\sim$20\,MeV before.  For heavy lepton neutrinos the energy is even higher, $\sim$50\,MeV (compared to $\sim$25\,MeV before black hole formation).  The reason for the increase in the average energy is the strong energy dependence of the coherent scattering which leads to the high energy neutrinos being preferentially scattered (and therefore their arrival delayed) in the accretion flow. 

For canonical failed supernovae, at Galactic distances of 10\,kpc, we predict that current and near-future detectors will see $\sim$2 events per 10\,kT of target material.  For Super-K, a 32\,kT water Cherenkov detector or JUNO, a 20\,kT liquid scintillator detector, we predict $\sim$5 neutrino-echo neutrinos, while for DUNE, a 40\,kT liquid Argon detector we expect $\sim$9 neutrinos.  For Hyper-K, a 220\,kT water Cherenkov detector, the expected number of neutrino-echo neutrinos is $\sim$35. These numbers assume adiabatic MSW oscillations and the normal mass ordering of neutrinos, however there is not a significant change if one assumes the inverted mass ordering.  While JUNO, DUNE, and Hyper-K are future detectors, while Super-K and IceCube are currently running, making a neutrino-echo neutrino detection possible today with a Galactic failed supernova.  Due to the large background in the IceCube detector, the total number of neutrino-echo events is not so interesting, rather what is of interest is how much the signal rises above the noise level.  We find for the model explored here, and for a failed supernova at 10\,kpc, that a signature of the neutrino echo should be detectable in IceCube as a 3-5$\sigma$ excess in the background noise in the first few milliseconds directly following the main signal cutoff from the black hole formation.

For all but the lowest neutrino energies, the neutrino echo quickly overtakes the expected exponential decay of the neutrino signal coming from the protoneutron star as it accretes into the black hole. Nevertheless, we confirm with our detailed physics simulations that the decline of these low energy neutrinos reaches and is maintained at the analytic value of $\sqrt{27}M$.

The strength of the neutrino echo depends on the properties of the main neutrino emission in the final 10s of milliseconds before black hole formation.  This should be well measured with neutrino detectors for a galactic supernova.  The remaining dependence in the strength of the echo is from the matter properties in the accretion flow, namely the composition, which sets the coherent scattering cross section, and the density.  Therefore an observation or limit on neutrino-echo neutrinos can be used to constrain this flow.  Key future work will be determining the neutrino echo signal for multidimensional simulations, where the matter structure and composition surrounding the protoneutron star may differ from the spherically symmetric model explored here.

\begin{acknowledgments}
We thank Sherwood Richers for discussions surrounding SedonuGR and Shuai Zha for discussions throughout the course of this work. This work is supported by the Swedish Research Council (Project No. 2020-00452)
and the Science and Technology Facilities Council (United Kingdom).
The simulations were enabled by resources provided by the Swedish National Infrastructure for Computing (SNIC) at PDC and NSC partially funded by the Swedish Research Council through grant agreement No. 2016-07213.

\end{acknowledgments}

\software{GR1D \citep{oconnor:10,oconnor:15}, NuLib \citep{oconnor:15},  
          SedonuGR \citep{richers:15,richers:17}, 
          SNEWPY \citep{snewpy}, SNOwGLoBES \citep{snowglobes}, matplotlib \citep{Hunter:2007}
          }

\end{document}